\documentclass[showpacs,preprintnumbers,amsmath,amssymb,floatfix]{revtex4}

\usepackage{epsfig}
\usepackage{graphicx}
\usepackage{dcolumn}
\usepackage{bm}
\usepackage{amsmath}
\usepackage{slashbox}
\usepackage{multirow}
\usepackage{color}
\usepackage{mathrsfs}

\begin{document}

\def\be{\begin{equation}}
\def\ee{\end{equation}}
\def\bea{\begin{eqnarray}}
\def\eea{\end{eqnarray}}
\newcounter{Lcount}
\def\bl{\setcounter{Lcount}{0}
\begin{list}{\arabic{Lcount}.}{\usecounter{Lcount}\setlength{\leftmargin}{0.4cm}}}
\def\el{\end{list}}

\title{Fully double-logarithm-resummed cross sections}
\author{S.\ Albino}
\affiliation{{II.} Institut f\"ur Theoretische Physik, Universit\"at Hamburg,\\
             Luruper Chaussee 149, 22761 Hamburg, Germany}
\author{P.\ Bolzoni}
\affiliation{{II.} Institut f\"ur Theoretische Physik, Universit\"at Hamburg,\\
             Luruper Chaussee 149, 22761 Hamburg, Germany}
\author{B.\ A.\ Kniehl}
\affiliation{{II.} Institut f\"ur Theoretische Physik, Universit\"at Hamburg,\\
             Luruper Chaussee 149, 22761 Hamburg, Germany}
\author{A.\ Kotikov}
\affiliation{{II.} Institut f\"ur Theoretische Physik, Universit\"at Hamburg,\\
             Luruper Chaussee 149, 22761 Hamburg, Germany}
\affiliation{Bogoliubov Laboratory of Theoretical Physics, JINR, 141980 Dubna, Russia}
\date{\today}
\begin{abstract}
We calculate the complete double logarithmic contribution to cross sections for semi-inclusive  
hadron production in the modified minimal-subtraction ($\overline{\rm MS}$) scheme by applying dimensional regularization to the double logarithm approximation.
The full double logarithmic contribution to the coefficient functions for inclusive hadron production in $e^+ e^-$ annihilation
is obtained in this scheme for the first time.
Our result agrees with all fixed order results in the literature, which extend to next-next-to-leading order.
\end{abstract}


\maketitle


\section{Introduction}
\label{Intro}

The inclusive production of particles in the framework of the factorization theorem and perturbation theory 
at high energy has been understood since a long time.
However, perturbation theory fails when the fraction $x$ of available energy carried away by the observed particle is too low.
Specifically, large unresummed logarithms spoil the convergence of the series.
The wealth of data at lower $x$ that have to be excluded from global fits pending
the explicit resummation of such logarithms would significantly improve the constraints on fragmentation functions (FFs)
at large $x$ and, for the first time, at small $x$,
as well as on $\alpha_s(M_Z)$.

In fact, the largest logarithms, the {\it double logarithms} (DLs), in the splitting functions that determine the evolution of the
FFs have been determined to all orders a long time ago \cite{Mueller:1981ex}, and have even been used to perform 
leading order (LO) global fits 
\cite{Albino:2005gd,Albino:2005gg} to data measured at the smallest $x$ values.
Specifically, in these fits the evolution was calculated in the fixed order (FO) approach, to allow 
for a description of the large $x$ data, while including the complete DL contribution to all orders, and this
consistent approach resulted in a {\it simultaneously} good description of the remaining smaller $x$ data.

Strictly speaking, the theoretical approach used at LO in the global fits of Refs.\ \cite{Albino:2005gd,Albino:2005gg}
to measurements of inclusive particle production in $e^+ e^-$ collisions
is incomplete,
because the DLs in the coefficient functions are not resummed.
These DLs are expected to be not as important as those appearing in the evolution because they only appear at and beyond NLO and, furthermore,
the approach turned out to be adequate for the numerical analysis.
However, the inclusion of the DLs in the coefficient functions could make a significant improvement to the 
accuracy of cross section calculations making the analysis of Refs.\ \cite{Albino:2005gd,Albino:2005gg}
feasible also at next-to-leading order (NLO).
The complete DL contribution to partonic cross sections has been calculated 
\cite{Mueller:1982cq} for the case in which the collinear singularities are regularized by giving a 
small mass $m_g$ to the gluon, the so-called massive gluon (MG) regularization scheme. 
This result turns out to be inconsistent with the full next-next-to-leading order (NNLO) result in the modified minimal-subtraction ($\overline{\rm MS}$) scheme.
In the literature, the $O(\alpha_s^2)$ contributions to the fragmentation 
functions in $e^+ e^-$ annihilation have been originally computed 
in Refs.\ \cite{Rijken:1996vr,Rijken:1996npa,Rijken:1996ns}. 
Their results were confirmed 10 years later
using rather different technologies by the new independent computation in Mellin ($N$) space \cite{Mitov:2006wy}.
The Mellin space result of Ref.\ \cite{Mitov:2006wy}
agrees with the Mellin transform of the $x$-space result of Refs.\ \cite{Rijken:1996vr,Rijken:1996npa,Rijken:1996ns}
performed in \cite{Blumlein:2006rr}.
As a result of these last computations, several typographical errors present in the original papers
\cite{Rijken:1996vr,Rijken:1996npa,Rijken:1996ns} have been corrected. 
The inconsistency between the NNLO DLs calculated in Ref.\ \cite{Mueller:1982cq} and 
those calculated in Refs.\ \cite{Rijken:1996vr,Rijken:1996npa,Rijken:1996ns,Mitov:2006ic,Mitov:2006wy}
is not surprising --- the two computations
were carried out in two different regularization and factorization schemes. 

Therefore, it is necessary to calculate the DL contribution to cross sections also in 
the $\overline{\rm MS}$ scheme, which is the goal of this paper.
In section \ref{DLswithmomcutoff}, we review the derivation of the double logarithm approximation (DLA) result
calculated in the MG scheme, and its factorization, for the probability of the inclusive
production of an observed gluon. Indeed, it is the gluon channel which contains the DLs.
In section \ref{DLswithdimreg}, we calculate the modification to this result when dimensional regularization (DR)
is used instead,
and factorize it in the $\overline{\rm MS}$ factorization scheme to obtain the complete DL
contribution to the splitting functions and coefficient functions in this scheme.
Finally, in section \ref{conc} we present our conclusions.

\section{Double logarithms in the massive gluon scheme \label{DLswithmomcutoff}}

We begin by repeating the derivation of the result for the DL contributions to splitting functions and coefficient functions
of Ref.\ \cite{Mueller:1982cq} in a way which makes suitable its comparison with our derivation using 
dimensionally regularization, to be presented in section \ref{DLswithdimreg}. In this section we work in the
MG scheme in order to regularize the mass singularities, which is why we assign a small mass $m_g$ to each
gluon.

As is well known, these DL contributions appear in the gluon-gluon and gluon-quark
timelike splitting functions \cite{Mueller:1981ex} and in the timelike gluon coefficient function
\cite{Mueller:1982cq}. To extract them we
consider a general process with a colour singlet final state
involving the production of an ``observed'' gluon of momentum $q$ from a hard 
parton of momentum $p$ around which a jet is formed.
In the DLA, the DL contribution arises from unobserved soft gluons in the final state.
Consequently, there must be an additional hard parton to account for the recoil from the parton of momentum $p$ as a result 
of momentum conservation. The momentum of this additional parton will be written $\bar{p}$.
The cross section for this process will be written $d\sigma(p,\bar{p},q)$.
A typical example is the process
$e^+ + e^- \rightarrow V^* \rightarrow {\mathcal Q}(p) +\bar{\mathcal Q}(\bar{p}) +g(q) +X $, 
where $V^*=\gamma, Z$ is a virtual vector boson, where the jet is formed around the quark ${\mathcal Q}$
with momentum $p$ and around the antiquark $\bar{\mathcal Q}$ with momentum $\bar{p}$, and where
$X$ is any hadronic final state that is allowed by quantum number conservation. 
Thus, to obtain the DL contribution to the cross section, we consider the configuration in which 
the unobserved part consists of only 
$N$ soft gluons of momenta $q_1,q_2,\ldots,q_N$ whose phase space is fully integrated out.  
Therefore, defining $d\sigma_N(p,\bar{p},k_1,k_2,\ldots,k_N)$ to be the cross section 
in which $N$ gluons of momenta $k_\alpha$, $\alpha=1,2,\dots,N$, are produced together with the partons of momentum $p$ and $\bar{p}$, we can write
\be
d\sigma(p,\bar{p},q)=\sum_{N=0}^\infty d\sigma_{N+1}(p,\bar{p},q,q_1,q_2,\ldots,q_N),
\label{dsigppprimeq}
\ee
where it is understood that the $q_\alpha$ are fully integrated over, but not $q$.
It is a well known result \cite{Bassetto:1982ma} that the DL
contributions come from the kinematic configuration in which the momenta
of the soft gluons are strongly ordered, i.e.
\be
|\vec{q}|\ll |\vec{q_1}|\ll |\vec{q_2}|\ll \cdots \ll |\vec{q_N}| \ll Q/2, 
\label{momord}
\ee 
where $Q^2$ is the perturbative scale, i.e.\ the scale that the factorization and renormalization scales should have the same
order of magnitude as. For a general process, the choice $Q^2 =(p+\bar{p})^2$ is suitable and is usually made, in particular this choice is made
in $e^+ e^-$ annihilation. 
Furthermore, as initially explicitly
proved up to three loops \cite{Mueller:1981ex} and then formally at all orders in
Ref.\ \cite{Bassetto:1982ma}, to extract the most singular terms 
the phase space should also be restricted to the region where the angles $\theta_I$ of the emitted
gluons with respect to the hard parton of momentum $p$ are strongly ordered, i.e.\
\be
\theta \ll \theta_1\ll\theta_2\ll\cdots\ll\theta_N\ll 1,
\label{angord}
\ee   
where $\theta$ refers to the gluon of momentum $q$. 
Now, in Appendix \ref{appendix}, we derive the factorization of soft gluon emission without carrying out the phase space integrations
for any of the soft gluons. 
Thus,
using Eq.\ (\ref{dsigNp1fromdsigN2}), which is the general result for the factorization of soft gluon emission, 
but choosing $N+1$ final state gluons and setting $k_1=q$ and $k_{\alpha+1} = q_\alpha$ for $\alpha=1,2,\dots,N$,
$d\sigma_{N+1}$ at DL accuracy factorizes in the following way:
\be
d\sigma_{N+1}(p,\bar{p},q,q_1,q_2,\ldots,q_N)=d\sigma_B(p,\bar{p})dw_g(q)dw_g(q_1) dw_g(q_2) \ldots  dw_g(q_{N-1})dw_I(q_N),
\label{dsigNp1fromdsigN}
\ee
where $d\sigma_B(p,\bar{p})$ is the Born cross section of the underlying process --- 
e.g.\ in $e^+ e^-$ annihilation above it is the LO cross section 
for the process $e^+ e^-\rightarrow {\mathcal Q}(p)+\bar{\mathcal Q}(\bar{p})$,
and $I$ labels the species of the initial state parton with momenta $p$ (see Appendix \ref{appendix}).
According to Eq.\ (\ref{singfacth}), $dw_I(q_\alpha)$ is, in $d=4$ dimensions,
given by:
\be
dw_I(q_\alpha)=g^2\, K_J 
\, \frac{2 (p \cdot \bar{p})}{(p \cdot q_\alpha) (\bar{p} \cdot q_\alpha)}\,
\frac{d^{3}q_\alpha}{(2\pi)^{3}2q_\alpha^0},
\label{weight4}
\ee
where $g$ is the strong coupling constant and $K_J$ is defined in Appendix \ref{appendix}.
We will work in a frame where $\bar{\bf p}=-{\bf p}$. Aligning the $z$-axis with ${\bf p}$, this means that
\be
p =\frac{Q}{2}(1,{\bf{0}}_\perp,1),\qquad 
\bar{p} =\frac{Q}{2}(1,{\bf{0}}_\perp,-1).
\ee
We introduce the usual Sudakov parametrization for the gluon momenta $q_\alpha$, i.e.
\be
q_\alpha=x_\alpha\, p + z_\alpha x_\alpha \,\bar{p} + q_{\alpha\perp},
\qquad 
q_{\alpha\perp}=(0,{\bf{q}}_{\alpha\perp},0),
\qquad 
{\bf{q}}_{\alpha\perp}^2=z_\alpha x_\alpha^2\, Q^2 - m_g^2,
\label{sudakov}
\ee
where we have used the on-shell condition for the gluon in the MG regularization scheme
(i.e.\ $q_\alpha^2=m_g^2$). From Eq.\ (\ref{sudakov}), we find that the quantities $x_\alpha$ and
$z_\alpha$ are given by
\begin{eqnarray}
z_\alpha&=&\frac{p \cdot q_\alpha}
{\bar{p} \cdot
q_\alpha}=\frac{1-|\vec{q}_\alpha|/q_\alpha^0\cos\theta_\alpha}{1+|\vec{q}_\alpha|/q_\alpha^0\cos\theta_\alpha},
\label{trans1}
\\
x_\alpha&=&\frac{\bar{p} \cdot q_\alpha}{\bar{p} \cdot p}=\frac{2q^0_\alpha}{Q(1+z_\alpha)},
\label{trans2}\\
q_\alpha^0&=&\sqrt{|\vec{q}_\alpha|^2+m_g^2}.
\label{onshellcond}
\end{eqnarray}
We can now use Eqs.\ (\ref{trans1}) and (\ref{trans2}) to change variables in Eq.\ (\ref{weight4}) from $q_\alpha$, $p$ and $\bar{p}$ to
$x_\alpha$ and $z_\alpha$.
Calculating the Jacobian determinant of this change of variables and using 
$d^3 q_\alpha=2\pi |\vec{q}_\alpha|^2d|\vec{q}_\alpha|d\cos\theta$, where the factor $2\pi$ arises from the azimuthal integration
using the symmetry of $dw_I$ around the $z$-axis, 
the result is that
\be
dw_I(x_\alpha,z_\alpha)=2a_s K_J \frac{dx_\alpha}{x_\alpha}\, \frac{dz_\alpha}{z_\alpha}+O(m_g^2),
\label{weight4xz}
\ee
where $a_s=\alpha_s/(2\pi)=g^2/8\pi^2$ is the bare coupling, which at this level 
of accuracy can be replaced by the renormalized coupling. In Eq.\ (\ref{weight4xz}), we have neglected
terms proportional to $m_g^2$ which do not affect the DLs we are interested in.
However, in the MG regularization scheme, the small gluon mass $m_g$ results in a non-zero lower 
cutoff for 
the variables $z_\alpha$ appearing in the denominator of Eq.\ (\ref{weight4xz}). 
To find the value of this small cutoff $Z_\alpha$,
we note from Eq.\ (\ref{trans1}) that the minimum of $z_\alpha$ corresponds to the 
collinear limit, i.e.\ $\theta_\alpha = 0$. In this limit,
using Eq.\ (\ref{trans2}) together with the on-shell condition in Eq.\ (\ref{onshellcond}) 
to express $|\vec{q}_\alpha|/q_\alpha^0$ in terms of
$x,m_g$ and $z$,
Eq.\ (\ref{trans1}) leads to the following implicit equation for the cutoff $Z_\alpha$:
\be
Z_\alpha-\frac{1-\sqrt{1-\frac{4 m_g^2}{x_\alpha^2 Q^2(1+Z_\alpha)^2}}}{1+\sqrt{1-\frac{4 m_g^2}{x_\alpha^2 Q^2(1+Z_\alpha)^2}}}
=0.\ee
This has the exact solution
\be
Z_\alpha=\frac{m_g^2}{x_\alpha^2 Q^2}.
\label{cutoff}
\ee
The physical meaning of the variables $x_\alpha$ and $z_\alpha$ defined in 
Eqs.\ (\ref{trans1}) and (\ref{trans2}) becomes clearer if we keep the first order
term in the small gluon mass $m_g$ and in the small angles $\theta_\alpha$
as required by Eq.\ (\ref{angord}):
Defining $\omega_\alpha=q_\alpha^0$,
\be
x_\alpha=\frac{\omega_\alpha}{p_0}\left[1+O\left(\frac{m_g^2}{\omega_\alpha^2}\right)+O\left(\theta_\alpha^2\right)\right],\qquad 
z_\alpha=\frac{\theta_\alpha^2}{4}+O\left(\frac{m_g^2}{\omega_\alpha^2}\right)+O\left(\theta_\alpha^4\right).\label{approxdefofx}
\ee
It is clear that if we use these expressions to define $x_\alpha$ and $z_\alpha$ appearing
in Eq.\ (\ref{weight4xz}), together with the 
collinear cutoff in Eq.\ (\ref{cutoff}) to compute the DLs according to
Eqs.\ (\ref{dsigppprimeq}) and (\ref{dsigNp1fromdsigN}), only $O(m_g^2)$ terms are modified, 
but not the terms which are singular or just non-vanishing
as $m_g \rightarrow 0$.
We may therefore set $x_0=x$, where $x$ is the usual energy or momentum fraction, $z_0=z$ and $Z_0=Z$.
Since $x$ is measured in experiment but the remaining degrees of freedom in $q$ are not (i.e.\ $z$ is integrated over), 
we replace $dw_g(x,z)$ with $dw_g(x',z) \delta(x-x')dx$
on the right hand side of Eq.\ (\ref{dsigppprimeq}) and then divide both sides by $dx$, giving 
\be
\frac{d\sigma}{dx}(p,\bar{p},x,Z)=d\sigma_B(p,\bar{p}) \sum_{N=0}^\infty \delta(x'-x)dw_I(x',z_0) dw_g(x_1,z_1)dw_g(x_2,z_2)  \ldots
dw_g(x_N,z_N).
\label{dsigppprimeq2}
\ee
In Eq.\ (\ref{dsigppprimeq2}), we have noted that there is
an explicit dependence on $m_g$, which determines the lower bound in the integrations over $z_\alpha$
according to Eq.\ (\ref{cutoff}).

Finally, putting Eq.\ (\ref{weight4xz}) into Eq.\ (\ref{dsigppprimeq2}), then using 
Eqs.\ (\ref{momord}) and (\ref{angord}) to explicitly determine the integration limits in terms of
various variables defined above including the collinear cutoff defined in Eq.\ (\ref{cutoff}),
we find that
\be
xg(x,Z)=2 a_s K_I\int_x^1 \frac{dx_1}{x_1}\int_Z^1 
\frac{dz^{\prime}}{z^{\prime}} x_1 G(x_1,z'),
\label{integrelbetweengandG}
\ee
where
\be
g(x,Z)=\frac{1}{d\sigma_B(p,\bar{p})}\frac{d\sigma}{dx}(p,\bar{p},x,Z)
\label{gfunction}
\ee
is the process-dependent (i.e.\ $I$-dependent) 
probability density in $x$ for the inclusive emission from an initial state parton of species $I$ of an observed gluon carrying away 
a momentum fraction $x$, and
\be
xG(x,Z)=\delta(1-x)+\sum_{N=1}^\infty (2 a_s C_A)^N \int_x^1 \frac{dx_2}{x_2}\int_{x_2}^1
\frac{dx_3}{x_3}\ldots \int_{x_{N-1}}^1 \frac{dx_N}{x_N}
\int_Z^1 \frac{dz_1}{z_1}\int_{z_1}^1 \frac{dz_2}{z_2}\ldots \int_{z_{N-1}}^1
\frac{dz_N}{z_N}
\label{tradresforG}
\ee
is the associated process-independent probability for the inclusive emission from an initial state
single gluon of an observed gluon carrying away 
a momentum fraction $x$ and emitted at an angle $\nu$ obeying $Z=(1-\cos\nu)/2$.
Notice that Eq.\ (\ref{tradresforG}) explicitly contains the conditions
$z_\alpha > z_{\alpha-1}$ for $\alpha=2,3,\dots,N$ dictated by the angular
ordering in Eq.\ (\ref{angord}) plus the condition $z_1>Z$.
However, it is also compatible with the less restrictive conditions
$z_\alpha > Z_\alpha$ for $\alpha=1,2,\dots,N$, as may be understood by also
exploiting the conditions $Z_{\alpha-1} > Z_\alpha$ for $\alpha=2,3,\dots,N$
dictated by the momentum ordering in Eq.\ (\ref{angord}).
In fact, we have $z_1>Z>Z_1$ and $z_\alpha > z_{\alpha-1} > Z_{\alpha-1} > Z_\alpha$
for $\alpha=2,3,\dots,N$, so that it follows by iteration that
$z_\alpha > Z_\alpha$ is satisfied for all $\alpha=1,2,\dots,N$.
We note that, in Eq.\ (\ref{integrelbetweengandG}), 
$z^{\prime}=m_g^2/(x^{\prime 2} Q^{\prime 2})$ plays the role of the collinear cutoff 
in the calculation of $G$ in Eq.\ (\ref{tradresforG}),
where $Q^{\prime 2}=z x^2 Q^2$ is the energy scale relevant to $G$ 
in the same way as $Q$ is the energy scale relevant to $g$
because, using Eq.\ (\ref{sudakov}), we see that $Q^{\prime 2}$ is the transverse energy of the first gluon emitted
from the initial state parton of momentum $p$.

In order to factorize $g$, we will first consider the factorization of $G$, 
which turns out to be simpler and more fundamental. The quantity
$g$, which can be more directly related to physical observables (see later),
can then be obtained from $G$ via
\be
g(x,Z)=\frac{K_I}{C_A}\left[G(x,Z)-\delta(1-x)\right].
\label{gxfromGx}
\ee
We observe that this equation, together with the definition of $g(x,Z)$ in
Eq.\ (\ref{integrelbetweengandG}), represents the integral master equation
for $G(x,Z)$ originally derived in the framework of a pure gluonic
theory in Ref.\ \cite{Mueller:1982cq}, whose solution is actually given by Eq.\ (\ref{tradresforG}). 

Performing the integrals in Eq.\ (\ref{tradresforG}) gives
\be
G\left(x,Z\right)
=\delta(1-x)+\frac{1}{x}\sum_{N=1}^\infty (2C_A a_s)^N \frac{1}{(N-1)!} \ln^{N-1}\frac{1}{x}\frac{1}{N!}\ln^N \frac{1}{Z}.
\label{Gexplicit}
\ee
We define
\be
{\mathbf G}\left(x,\frac{m_g^2}{Q^2}\right)=G\left(x,Z(x)\right).
\ee
In Mellin space, defined for any function $f(x)$ via
\be
f(\omega)=\int_0^1 dx x^\omega f(x),
\ee
Eq.\ (\ref{Gexplicit}) reads
\be
{\mathbf G}\left(\omega,\frac{m_g^2}{Q^2}\right)=\sum_{N=0}^\infty (2C_A a_s)^N 
\sum_{m=0}^N\frac{(-2)^m (N+m-1)! \ln^{N-m}\frac{Q^2}{m_g^2}}
{(N-1)!m!(N-m)!\omega^{N+m}},
\label{Ginomegaexpl}
\ee
where we have used $Z=m_g^2/x^2Q^2$ and the fact that
\be
\frac{1}{(r-1)!}\int_0^1 dx x^\omega \frac{\ln^{r-1}\frac{1}{x}}{x}=\frac{1}{\omega^{r}}.
\label{mellintrans}
\ee
Now, according to the QCD factorization theorem \cite{Ellis:1978ty,Curci:1980uw},
we know that all the collinear singularities in Eq.\ (\ref{Ginomegaexpl})
(which here appear as logarithms of the regulator $m_g$) can be factorized into a
transition function $\Gamma$ according to
\be
{\mathbf G}\left(\omega,\frac{m_g^2}{Q^2}\right)=C\left(\omega,a_s,\frac{Q^2}{\mu_F^2}\right)
\Gamma\left(\omega,a_s,\frac{\mu_F^2}{m_g^2}\right),
\label{factofG}
\ee
where 
\be
\Gamma\left(\omega,a_s,\frac{\mu_F^2}{m_g^2}\right)=
\exp\left[\gamma(\omega,a_s)\ln \frac{\mu_F^2}{m_g^2}\right],
\ee
with $\mu_F^2$ being an arbitrary factorization scale.
Choosing $\mu_F^2=Q^2$, writing $C(\omega,a_s,1)=C(\omega,a_s)$
and expanding Eq.\ (\ref{factofG}) in powers of $\ln (Q^2/m_g^2)$ gives
\be
{\mathbf G}\left(\omega,\frac{m_g^2}{Q^2}\right)=C(\omega,a_s)-\gamma(\omega,a_s)C(\omega,a_s)
\ln\frac{m_g^2}{Q^2}+O(\ln^2(m_g^2/Q^2)).
\label{logexp}
\ee  
Therefore, comparing Eq.\ (\ref{logexp}) with 
the part of the summation in Eq.\ (\ref{Ginomegaexpl}) for which $m=N$ (i.e.\ the coefficient of $\ln^0\frac{m_g^2}{Q^2}$) 
and then for which $m=N-1$ (i.e.\ the coefficient of $\ln^1\frac{m_g^2}{Q^2}$),
we easily obtain that 
\be
C(\omega,a_s)=\sum_{N=0}^\infty (2C_A a_s)^N 
\frac{(-2)^N (2N-1)!}
{(N-1)!N!\omega^{2N}}.
\ee
and
\be
\gamma(\omega,a_s)C(\omega,a_s)=\sum_{N=0}^\infty (2C_A a_s)^N 
\frac{(-2)^{N-1} (2N-2)!}
{[(N-1)!]^2\omega^{2N-1}}.
\label{gammac}
\ee
These sums may be respectively identified to obtain, for the coefficient function,
\be
C(\omega,a_s)=\frac{1}{2}\frac{\omega+\sqrt{\omega^2+16C_A
a_s}}
{\sqrt{\omega^2+16C_A a_s}},
\label{coefffuncmg}
\ee
as well as the result determining the anomalous dimension,
\be
\gamma(\omega,a_s)C(\omega,a_s)=\frac{2 C_A a_s}{\sqrt{\omega^2+16C_A a_s}},
\ee
i.e.\
\be
\gamma(\omega,a_s)=\frac{1}{4}\left(-\omega+\sqrt{\omega^2+16C_Aa_s}\right).
\label{finresforgamma}
\ee

The quantity $\Gamma$ obeys the timelike DGLAP equation,
\be
\frac{d\Gamma}{d\ln \mu_F^2}=\gamma \Gamma,
\label{DGLAPforGamma}
\ee
so that the dependence of the coefficient function on the factorization scale is
easily obtained by imposing the independence on $\mu_F^2$ of the gluon density function $G$
in Eq.\ (\ref{factofG}) at the same level of accuracy.

Next we turn our attention to the factorization of $g$.
Defining
\be
{\mathbf g}\left(x,\frac{m_g^2}{Q^2}\right)=g(x,Z),
\ee
Eq.\ (\ref{gxfromGx}) in Mellin space becomes
\be
{\mathbf g}\left(\omega,\frac{m_g^2}{Q^2}\right)=\frac{K_I}{C_A}\left[{\mathbf G}\left(\omega,\frac{m_g^2}{Q^2}\right)-1\right].
\ee
The general factorization formula reads
\be
(f,{\mathbf g})=(C_q,C_g)\hat{\Gamma},
\label{factoffg}
\ee
where $f$ describes the equivalent partonic process to $g$ but for the inclusive production of a quark instead of a gluon,
$\hat{\Gamma}$ is a 2$\times$2 matrix containing all collinear singularities in $(Q,g)$,
and $C_q$ and $C_g$ are respectively the collinear-singularity-free quark and gluon coefficient functions.
In the DLA, $f=C_q=1$ so that, from Eqs.\ (\ref{factofG}) and (\ref{factoffg}), we find that
\be
C_g\left(\omega,a_s,\frac{Q^2}{\mu_F^2}\right)=\frac{K_I}{C_A}\left[C\left(\omega,a_s,\frac{Q^2}{\mu_F^2}\right)-1\right].
\label{CgfromC}
\ee
Thus, defining $C_g(\omega,a_s,1)=C_g(\omega,a_s)$, the resummed gluon coefficient function in the 
massive gluon regularization scheme is given by
\be
C_g(\omega,a_s)=\frac{K_I}{C_A}\frac{1}{2}\left[\frac{\omega}{\sqrt{\omega^2+16C_A a_s}}-1\right].
\label{gluoncoefff}
\ee
Furthermore, from Eqs.\ (\ref{factofG}) and (\ref{factoffg}), we find that
\begin{eqnarray}
\hat{\Gamma}=\left(
\begin{array}{cc}
1 & \frac{2C_F}{C_A}(\Gamma-1) \\ 
0 & \Gamma
\end{array}\right).
\label{hatgamfromgam}
\end{eqnarray}
Using Eq.\ (\ref{DGLAPforGamma}), we find that $\hat{\Gamma}$ obeys the DGLAP equation
\be
\frac{d}{d\ln \mu_F^2} \hat{\Gamma} = P^{\rm DL} \hat{\Gamma},
\label{DGLAPforGamhat}
\ee
where $P^{\rm DL}=\gamma A$ is the complete DL contribution to all the splitting functions, with
\begin{eqnarray}
A =\left(
\begin{array}{cc}
0 & \frac{2C_F}{C_A} \\
0 & 1
\end{array}\right).
\end{eqnarray}
Note that Eq.\ (\ref{hatgamfromgam}) can be obtained by solving Eq.\ (\ref{DGLAPforGamhat}) 
with the boundary condition $\hat{\Gamma}(\omega,a_s,1)=1$, and then using the projection operator property $A^2=A$, i.e.\
\be
\hat{\Gamma}\left(\omega,a_s,\frac{\mu_F^2}{m_g^2}\right)=\exp\left[P^{\rm DL}(\omega,a_s)\ln \frac{m_g^2}{\mu_F^2}\right]
=1+A\left(\exp\left[\gamma(\omega,a_s)\ln \frac{m_g^2}{\mu_F^2}\right]-1\right).
\ee

To check $C(\omega,a_s)$ and $\gamma(\omega,a_s)$ against full FO results in the literature, 
we consider the probability distribution $g$ for the inclusive production of a gluon in
$e^+ e^-$ annihilation. In this case, $K_I =2C_F$, where the factor 2 accounts for the fact that the
gluons can be emitted either collinearly to the quark of momentum $p$ or the quark of momentum $\bar{p}$.
Up to NNLO, $P^{\rm DL}$ agrees with the FO results for the DL contribution to the splitting functions 
calculated in the literature, as shown in Ref.\ \cite{Albino:2005gd}.
The NNLO expansion of Eq.\ (\ref{gluoncoefff}),
\be
C_g(\omega,a_s)=\frac{2C_F}{C_A}\left[-4C_A \frac{a_s}{\omega^2}+48 C_A^2 \left(\frac{a_s}{\omega^2}\right)^2\right],
\ee
agrees with Ref.~\cite{Mueller:1982cq}.
However, it
disagrees with the small $\omega$ limit of the full NNLO result e.g. in Ref.\ \cite{Blumlein:2006rr},
\be
C_g(\omega,a_s)=\frac{2C_F}{C_A}\left[-4C_A \frac{a_s}{\omega^2}+40 C_A^2 \left(\frac{a_s}{\omega^2}\right)^2\right].
\label{CgrestoNNLOinMSbar}
\ee
As already said, this discrepancy is not surprising, given that the result in Ref.\ \cite{Blumlein:2006rr}
was obtained using dimensional regularization followed by $\overline{\rm MS}$ factorization,
while here the result is obtained using a cutoff to regulate the collinear singularities. 
In section \ref{DLswithdimreg}, we will calculate the DL contribution in the $\overline{\rm MS}$ scheme, and show that
its NNLO expansion agrees with Eq.\ (\ref{CgrestoNNLOinMSbar}).

\section{Double logarithms in the $\overline{\rm MS}$ scheme \label{DLswithdimreg}}

In this section, we calculate the quantity $G$, this time using dimensional regularization. This will allow us to implement
$\overline{\rm MS}$ factorization. 
This requires deriving in terms of the variables $x_\alpha$ and $z_\alpha$ the probability $dw_I(q_\alpha)$
in $d\neq 4$ dimensions with a zero gluon mass, i.e.\ Eq.\ (\ref{eqfordwIinddim}) with $k=q_\alpha$. 
Now $d^{d-1}q_\alpha =|\vec{q}_\alpha|^{d-2} d|\vec{q}_\alpha| 
\sin^{d-4}\theta_\alpha \,d\cos\theta_\alpha \,d\Omega_{d-2}$,
where $\Omega_d$ is the $d$-dimensional solid angle. 
Because of the azimuthal symmetry of Eq.\ (\ref{eqfordwIinddim}) with respect to $\vec{p}$, 
the integrand is independent of $\Omega_{d-2}$, which may therefore be explicitly integrated out
using $\int d \Omega_d = d \pi^{d/2}/\Gamma(d/2+1)$
to give 
\be
dw_I(q_\alpha)=2 a_s \mu^{2\epsilon} \frac{(4\pi)^\epsilon}{\Gamma(1-\epsilon)}K_I \frac{2 (p\cdot \bar{p})}{(p\cdot q_\alpha) (\bar{p} \cdot q_\alpha)} 
|\vec{q}_\alpha|^{1-2\epsilon}\sin^{-2\epsilon}\theta_\alpha\, d|\vec{q}_\alpha| d\cos\theta_\alpha,
\ee
where $\mu$ is an arbitrary parameter with the dimension of mass (called the dimensional regularization mass)
which is needed to ensure that the action is dimensionless (note that $g\mu^\epsilon$ 
is independent of $\mu$ and $g$ is dimensionless).
Now, performing the change of variables given in Eqs.\ (\ref{trans1}) -- (\ref{onshellcond})
with $m_g=0$ gives the dimensionally regularized alternative to Eq.\ (\ref{weight4xz}), i.e.\
\be
dw_I(x_\alpha,z_\alpha)=2a_s \left(\frac{\mu^2}{Q^2}\right)^\epsilon \frac{(4\pi)^\epsilon}{\Gamma(1-\epsilon)} 
K_I\frac{dx_\alpha}{x_\alpha^{1+2\epsilon}} \frac{dz_\alpha}{z_\alpha^{1+\epsilon}}.
\label{expforprobdwD}
\ee
We have omitted a factor $(1-z)^{-\epsilon}=1+O(z)$, since doing so does not affect the DL contribution.
In this case, the collinear cutoff $Z$ can be set to zero, because  
the integrals for $\epsilon <0$ are well-defined even in the limit $Z\rightarrow 0$.
Hence, Eq.\ (\ref{tradresforG}) is replaced with
\be
\begin{split}
xG(x,\epsilon)=\delta(1-x)+&x^{-2\epsilon} \sum_{N=1}^\infty X^N
\int_x^1 \frac{dx_1}{x_1^{1+2\epsilon}}\int_{x_1}^1 \frac{dx_2}{x_2^{1+2\epsilon}}\ldots
\int_{x_{N-2}}^1 \frac{dx_{N-1}}{x_{N-1}^{1+2\epsilon}}\\
&\times \int_0^1 \frac{dz_1}{z_1^{1+\epsilon}}\int_{z_1}^1 \frac{dz_2}{z_2^{1+\epsilon}}
\ldots \int_{z_{N-1}}^1 \frac{dz_N}{z_N^{1+\epsilon}}, 
\label{startforG0}
\end{split}
\ee
where
\be
X=2C_A a_s \frac{(4\pi)^\epsilon}{\Gamma(1-\epsilon)}\left(\frac{\mu^2}{Q^2}\right)^\epsilon.
\ee
For comparison with Ref.\ \cite{Mueller:1982cq}, we note from Eq.\ (\ref{startforG0}) 
that $G(x,\epsilon)=G(x,Z=0,\epsilon)$, where $G(x,Z,\epsilon)$ is another quantity defined implicitly
by the
integral master equation 
\be
x^{1+2\epsilon}G(x,Z,\epsilon)=\delta(1-x)+ X\int_x^1 \frac{dx^{\prime}}{x^{\prime}}\int_Z^1 
\frac{dz^{\prime}}{(z^{\prime})^{1+\epsilon}}\, x' G(x',z',\epsilon).
\label{masteqinddim}
\ee
Equation (\ref{masteqinddim}) represents the
$d=4-2\epsilon$ generalization of the four-dimensional master equation given by Eq.\ (\ref{gxfromGx}) together with
Eq.\ (\ref{integrelbetweengandG}).

In the $\overline{\rm MS}$ renormalization scheme, the bare coupling $a_s$ is related to the
renormalized coupling $a_s(\mu_R^2,\epsilon)$ via $a_s S_\epsilon \left(\mu^2/\mu_R^2\right)^\epsilon 
= a_s(\mu_R^2,\epsilon) +O(a_s^2)$, where $S_\epsilon = e^{\epsilon(\ln 4\pi -\gamma_E)}$, so that
\be
X=2C_A  a_s(\mu_R^2,\epsilon) \left(\frac{\mu_R^2}{Q^2}\right)^\epsilon [1+O(\epsilon^2)] +O(a_s^2),
\label{Xfromas}
\ee
where the $O(a_s^2)$ and $O(\epsilon^2)$ corrections 
are neglected because they 
would only result in terms in our final
factorized quantities which are subleading
relative to the DLs that we are interested in here.
The quantity $\mu_R$ is a renormalization scale, of which $X$ is formally independent.
Using the symmetry of the $z$ integrals, we obtain
\be
f_N=\int_0^1 \frac{dz_1}{z_1^{1+\epsilon}}\int_{z_1}^1 \frac{dz_2}{z_2^{1+\epsilon}}
\ldots \int_{z_{N-1}}^1 \frac{dz_N}{z_N^{1+\epsilon}}=
\frac{1}{N!}\left(\int_0^1 \frac{dz}{z^{1+\epsilon}}\right)^N=\frac{(-1)^N}{N! \epsilon^N}.
\ee
Similarly, the $x$ integrals are given by
\be
g_N(x)=\int_x^1 \frac{dx_1}{x_1^{1+2\epsilon}}\int_{x_1}^1 \frac{dx_2}{x_2^{1+2\epsilon}}\ldots
\int_{x_{N-2}}^1 \frac{dx_{N-1}}{x_{N-1}^{1+2\epsilon}}=\frac{1}{(N-1)!}
\left(\int_x^1 \frac{dx_1}{x_1^{1+2\epsilon}}\right)^{N-1}=
\frac{(x^{-2\epsilon}-1)^{N-1}}{(N-1)!(2\epsilon)^{N-1}}.
\ee
Therefore,
\be
G(x,X,\epsilon)=\delta(1-x)+\frac{1}{x}\sum_{N=1}^\infty X^N
\frac{(-1)^N x^{-2\epsilon}(x^{-2\epsilon}-1)^{N-1}}{N!(N-1)!2^{N-1}\epsilon^{2N-1}}.
\label{Gagain}
\ee
To calculate the Mellin transform, we first define and then evaluate using integration by parts the quantity
\be
I_r=\int_0^1 dx x^{\omega-1}(x^{-2\epsilon}-1)^r
=\frac{2\epsilon r}{\omega}\int_0^1 dx x^{\omega-1}(x^{-2\epsilon}-1)^{r-1}
x^{-2\epsilon}
=\frac{2\epsilon r}{\omega}(I_r+I_{r-1}).
\label{mellinrec}
\ee
Thus, $I_r=I_{r-1}/[\omega/(2\epsilon r)-1]$. Using $I_0=1/\omega$, we find that
\be
I_r=\frac{1}{\omega}\frac{1}{\prod_{p=1}^r\left(\frac{\omega}{2\epsilon p}-1\right)}.
\ee
Using this result, we may evaluate the Mellin transform of the $x$-dependent part 
in Eq.\ (\ref{Gagain}) to obtain
\be
\int_0^1 dx x^{\omega-1} x^{-2\epsilon}(x^{-2\epsilon}-1)^{k-1}
=\frac{\omega}{2\epsilon k} I_k = \frac{1}{2\epsilon k}\frac{1}
{\prod_{p=1}^k\left(\frac{\omega}{2\epsilon p}-1\right)},
\ee
where in the first step we have used Eq.\ (\ref{mellinrec}).
Thus
\be
G(\omega,X,\epsilon)=\sum_{k=0}^\infty \left(-\frac{X}{2\epsilon^2}\right)^k \frac{1}{k!}
\frac{1}{\prod_{p=1}^k\left(\frac{\omega}{2\epsilon}-p\right)}.
\label{finformofG}
\ee
We note that
\be
G(\omega,X,\epsilon)= {}_0 F_1 \left(; \ 1-\frac{\omega}{2\epsilon};\, 
\frac{X}{2\epsilon^2}\right),
\ee
where $_p F_q(a_1,a_2,\ldots,a_p;\ b_1,b_2,\ldots,b_q;\ z)$ is the generalized hypergeometric function.
(In fact, ${}_0 F_1$ is related to the Bessel function $I_\nu$ via
${}_0 F_1 \left(; \nu +1;z^2/4\right)=[\Gamma(\nu+1)/(z/2)^\nu]I_\nu(z)$.)
Now, direct factorization of the collinear singularities as $\epsilon \rightarrow 0$ in Eq.\ (\ref{finformofG}) by expanding it in $\epsilon$
does not seem possible. 
However, from Eq.\ (\ref{finformofG}), it is easy to check that $G$ satisfies the following simple
differential equation:
\be
\frac{\ddot{G}-\frac{\omega}{2\epsilon}\dot{G}}{G}=\frac{X}{2\epsilon^2},
\label{diffeq1}
\ee
where we have defined
\be
\dot{f}(X)=X\frac{d f}{dX}.
\label{devdef}
\ee
Therefore, knowing from the factorization theorem 
that the factorization of the collinear singularities in Eq.\ (\ref{finformofG}) takes the form \cite{Curci:1980uw}
\be
G(\omega,X,\epsilon) 
=C\left(\omega,a_s(\mu_F^2,\epsilon),\frac{Q^2}{\mu_F^2},\epsilon\right)\Gamma\left(\omega,a_s(\mu_F^2,\epsilon),\epsilon\right),
\label{factXS}
\ee
where the transition function $\Gamma$
contains all the $\epsilon \rightarrow 0$ collinear singularities, we can solve Eq.\ (\ref{diffeq1}) to obtain
the timelike coefficient function $C$ using the fact that it is finite as $\epsilon \rightarrow 0$.
In the $\overline{\rm MS}$ factorization scheme, the timelike splitting function,
\be
\gamma(\omega,a_s(\mu_F^2,\epsilon))=\frac{d}{d\ln \mu_F^2}\ln \Gamma(\omega,a_s(\mu_F^2,\epsilon),\epsilon),
\ee
is explicitly independent of $\epsilon$. Using the boundary condition $\Gamma(\omega,0,\epsilon)=1$ for the
transition function, we have
\be
\Gamma(\omega,a_s(\mu_F^2,\epsilon),\epsilon)=
\exp\left[\int_0^{\mu_F^2} \frac{dq^2}{q^2} \gamma(\omega,a_s(q^2,\epsilon))\right].
\label{msbartransfunc}
\ee
Using the equation for the running coupling,
\be
\frac{da_s(\mu_R^2,\epsilon)}{d\ln \mu_R^2}=-\epsilon a_s(\mu_R^2,\epsilon)-\sum_{n=0}^\infty \beta_n a_s^{n+2}(\mu_R^2,\epsilon),
\label{runcoup}
\ee
and the fact that $a_s(0,\epsilon)=0$,
we can perform a change of integration variable in Eq.\ (\ref{msbartransfunc}) from $q^2$ to $a_s(q^2,\epsilon)$, which gives
\be
\Gamma(\omega,a_s,\epsilon)=\exp\left[\int_0^{a_s}\frac{d a_s}{-\epsilon a_s-\sum_{n=0}^\infty \beta_n a_s^{n+2}} \gamma(\omega,a_s)\right].
\label{Gamfromgamandas}
\ee
To our accuracy,
\be
\Gamma(\omega,a_s,\epsilon)=\exp\left[-\frac{1}{\epsilon}\int_0^{a_s}\frac{d a_s}{a_s}
\gamma(\omega,a_s)\right],
\label{Gamfromgamandas2}
\ee
because the inclusion of the terms of $O(a_s^2)$ and higher in Eq.\ (\ref{runcoup}) does not affect the 
DL contribution.
For simplicity, we will set $\mu_R=\mu_F=Q$ from now on, so that
\be
X=2C_A  a_s(Q^2,\epsilon)
\ee
and
\be
G(\omega,X,\epsilon) 
=C\left(\omega,a_s(Q^2,\epsilon),\epsilon\right)\Gamma\left(\omega,a_s(Q^2,\epsilon),\epsilon\right),
\label{factXS2}
\ee
where $C\left(\omega,a_s,\epsilon\right)=C\left(\omega,a_s,1,\epsilon\right)$.

The simplest way to obtain the timelike splitting function $\gamma$ and the timelike 
coefficient function $C$ is to compute the left hand side of the differential equation
in Eq.\ (\ref{diffeq1}) using Eqs.\ (\ref{factXS2}) and (\ref{Gamfromgamandas2}), and then to
compare the two inhomogeneous terms. In this way, we will find 
two solvable implicit equations which will allow us to find closed
expressions for the DL contributions to $\gamma$ and $C$.
So, differentiating Eq.\ (\ref{factXS2}) with respect to $X$
and using the result 
\be
\dot{\Gamma}=-\frac{\gamma}{\epsilon}\Gamma,
\ee
that follows from Eq.\ (\ref{Gamfromgamandas2}), we obtain in the notation of Eq.\ (\ref{devdef}) the results
\be
\dot{G}=\left(\dot{C}-\frac{\gamma}{\epsilon}C\right)\Gamma
\ee
and
\be
\ddot{G}=\left(\ddot{C}-2\frac{\gamma}{\epsilon}\dot{C}-\frac{\dot{\gamma}}{\epsilon}C+\frac{\gamma^2}{\epsilon^2}C\right)\Gamma,
\ee
so that
\be
\frac{\ddot{G}-\frac{\omega}{2\epsilon}\dot{G}}{G}=\frac{1}{\epsilon^2}\left(\gamma^2+\frac{\omega \gamma}{2}\right)
-\frac{1}{\epsilon}\left(\left[2\gamma +\frac{\omega}{2}\right]\frac{\dot{C}}{C}+\dot{\gamma}\right)+\frac{\ddot{C}}{C}.
\label{diffeqforGCandgamma}
\ee
Note that keeping the factor $K_\epsilon=1+O(\epsilon^2)$ in Eq.\ (\ref{Xfromas}) 
would result in an $O(\epsilon^0)$ change in this expression coming
from the first term of $O(\epsilon^{-2})$, thus $K_\epsilon$ can be safely excluded.
Now, comparing the coefficients of $\epsilon^{-2}$ and $\epsilon^{-1}$ on the right 
hand side of Eq.\ (\ref{diffeqforGCandgamma}) with those of Eq.\ (\ref{diffeq1})
and noting that $\gamma$ is explicitly independent of $\epsilon$ gives respectively
\be
\gamma^2+\frac{\omega\, \gamma}{2}-\frac{X}{2}=0,
\label{eqforgamma}
\ee
and
\be
\frac{\partial \ln C}{\partial \gamma}=-\frac{1}{2}\frac{1}{\gamma+\frac{\omega}{4}}.
\label{eqforC}
\ee
Equation (\ref{eqforgamma}) is the same quadratic equation for $\gamma$ as the one appearing 
in Ref.\ \cite{Mueller:1983js,Albino:2005gd},
whose unique solution compatible with perturbation theory is
Eq.\ (\ref{finresforgamma}).
Solving Eq.\ (\ref{eqforC}) for $C$ gives
\be
C=\frac{A(\omega)}{\sqrt{\gamma+\frac{\omega}{4}}},
\ee
where $A(\omega)$ is an unknown constant of integration. Using 
Eq.\ (\ref{finresforgamma}) for $\gamma$
and the condition $C(\omega,0)=1$ from perturbation theory,
we obtain the result $A(\omega)=\sqrt{\omega}/2$, so that, finally, we find
\be
C\left(\omega,a_s\right)=\frac{1}{\left(1+16C_A \frac{a_s}{\omega^2}\right)^{\frac{1}{4}}}.
\ee
From Eq.\ (\ref{CgfromC}), we obtain
\be
C_g\left(\omega,a_s\right)=\frac{K_I}{C_A}\left[\frac{1}
{\left(1+16C_A \frac{a_s}{\omega^2}\right)^\frac{1}{4}}-1\right].
\label{finalres}
\ee
This is the main result of this paper.
As $\omega\rightarrow 0$, $C_g$ is finite and is approximately given by
\be
C_g(\omega,a_s)\approx \frac{K_I}{2C_A(C_A a_s)^{\frac{1}{4}}} \sqrt{\omega}.
\ee

To obtain $C_g$ in $x$ space, we expand it in $a_s$ to obtain
\be
C_g\left(\omega,a_s\right)=\frac{K_I}{C_A}
\left[\left(1+\frac{16C_A a_s}{\omega^2}\right)^{-\frac{1}{4}}-1\right]
=\frac{K_I}{C_A}\sum_{r=1}^\infty \frac{(-1)^r\left(1/4\right)_r}{r!} 
\left(\frac{16C_A a_s}{\omega^2}\right)^r,
\ee
where $(a)_k=\Gamma(a+k)/\Gamma(a)$ are the Pochhammer symbols.
Then, using the result in Eq.\ (\ref{mellintrans}),
we obtain
\be
C_g\left(x,a_s\right)=\frac{K_I}{C_A}\frac{1}{x\ln \frac{1}{x}}
\sum_{r=1}^\infty \frac{(-1)^r}{(2r-1)!}\frac{\left(1/4\right)_r}{r!}
\left(16C_A a_s \ln^2 \frac{1}{x}\right)^r.
\label{coeffDLpre}
\ee
Identifying the sum gives the result for the DL contribution to the gluon coefficient function in $x$ space,
\be
C_g\left(x,a_s\right)=
-4K_I a_s \frac{\ln \frac{1}{x}}{x} 
\ _1 F_2 \left(\frac{5}{4};\ \frac{3}{2},2;\ -4C_A a_s \ln^2 x\right).
\label{coeffDL}
\ee

Finally, a gluon coefficient function which is valid from both large to small $x$ and formally consistent with
the FO and SGL resummed approaches is the DL-resummed NLO gluon coefficient function, 
given by
\be
C_g^{\rm{DL}+\rm{NLO}}\left(x,a_s\right)=C_g^{\rm{NLO}}\left(x,a_s\right)+C_g^{\rm{DL}}
\left(x,a_s\right)
-\frac{4\, K_I a_s \ln x}{x}, 
\label{coeffDLpNLO}
\ee
where $C_g^{\rm{DL}}$ is given by Eq.\ (\ref{coeffDL}), $C_g^{\rm{NLO}}$ by the NLO expansion of the FO result for $C_g$ (e.g.\ for
$e^+ e^-$ annihilation, see Eq.\ (\ref{coeffNLO}) below), and
the last term in Eq.\ (\ref{coeffDLpNLO})
prevents double counting of the DL at NLO.

In the case of $e^+ e^-$ annihilation, we have $K_I=2C_F$, so that
\be
C_g^{\rm{DL}}\left(\omega,a_s\right)=\frac{2C_F}{C_A}\left[\frac{1}
{\left(1+16C_A \frac{a_s}{\omega^2}\right)^\frac{1}{4}}-1\right].
\label{Cginepem}
\ee
In contrast to Eq.\ (\ref{gluoncoefff}),
Eq.\ (\ref{Cginepem}) now agrees with the FO results in the literature, 
i.e.\ it is equal to Eq.\ (\ref{CgrestoNNLOinMSbar})
when expanded to NNLO. More generally,
Eq.\ (\ref{finalres}) resums all the DLs of the gluon timelike coefficient function for $e^+ e^-$ annihilation
in the $\overline{\rm MS}$ scheme.
The gluon timelike coefficient function only starts to contribute at NLO in QCD. 
The contribution at NLO in $x$ space is given by \cite{Altarelli:1979kv,Baier:1979sp}
\be
C_g^{\rm{NLO}}\left(x,a_s\right)=a_s C_F \left\{2\frac{1+(1-x)^2}{x}\left[
\ln(1-x)+2\ln x\right]\right\}.
\label{coeffNLO}
\ee
\begin{figure}
\begin{center}
\begin{minipage}[b]{8cm}
\centering
\includegraphics[scale=0.4]{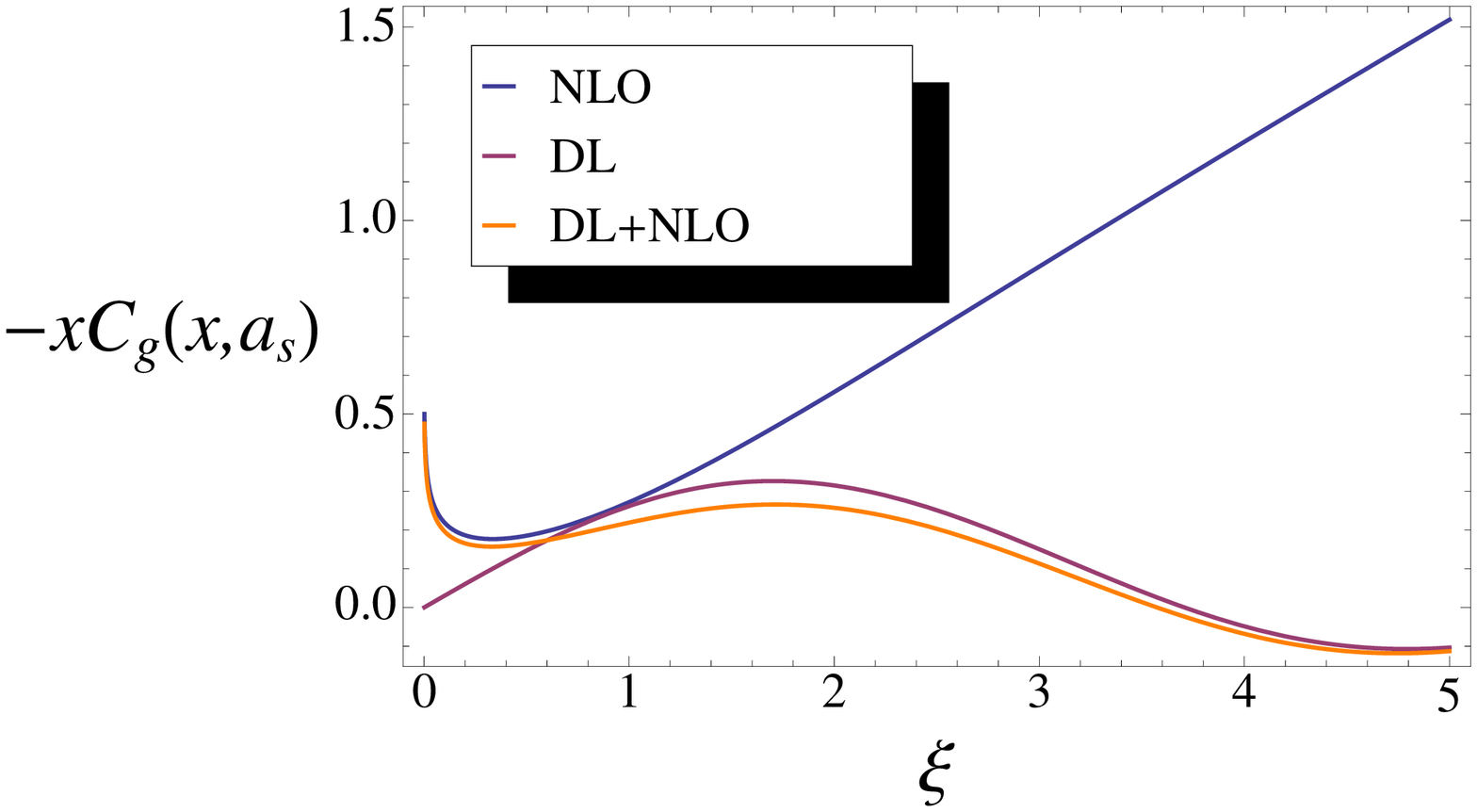}
\caption{\footnotesize{The gluon coefficient function in $e^+ e^-$ annihilation at NLO in Eq.\ (\ref{coeffNLO}) (blue lower line), 
the complete DL contribution to the gluon coefficient function in Eq.\ (\ref{coeffDLpre})} (violet line) and the DL-resummed 
NLO result in Eq.\ (\ref{coeffDLpNLO}) (orange line).
The value $a_s=0.18$ was used, because this corresponds to the result for $a_s(Q^2)$ at LO using
$Q=14$ GeV, the value for $Q$ used in Ref.\ \cite{Albino:2005gd}.}
\label{fig:coefficient}
\end{minipage}
\ \hspace{15mm}  
\begin{minipage}[b]{8cm}
\centering
\includegraphics[scale=0.4]{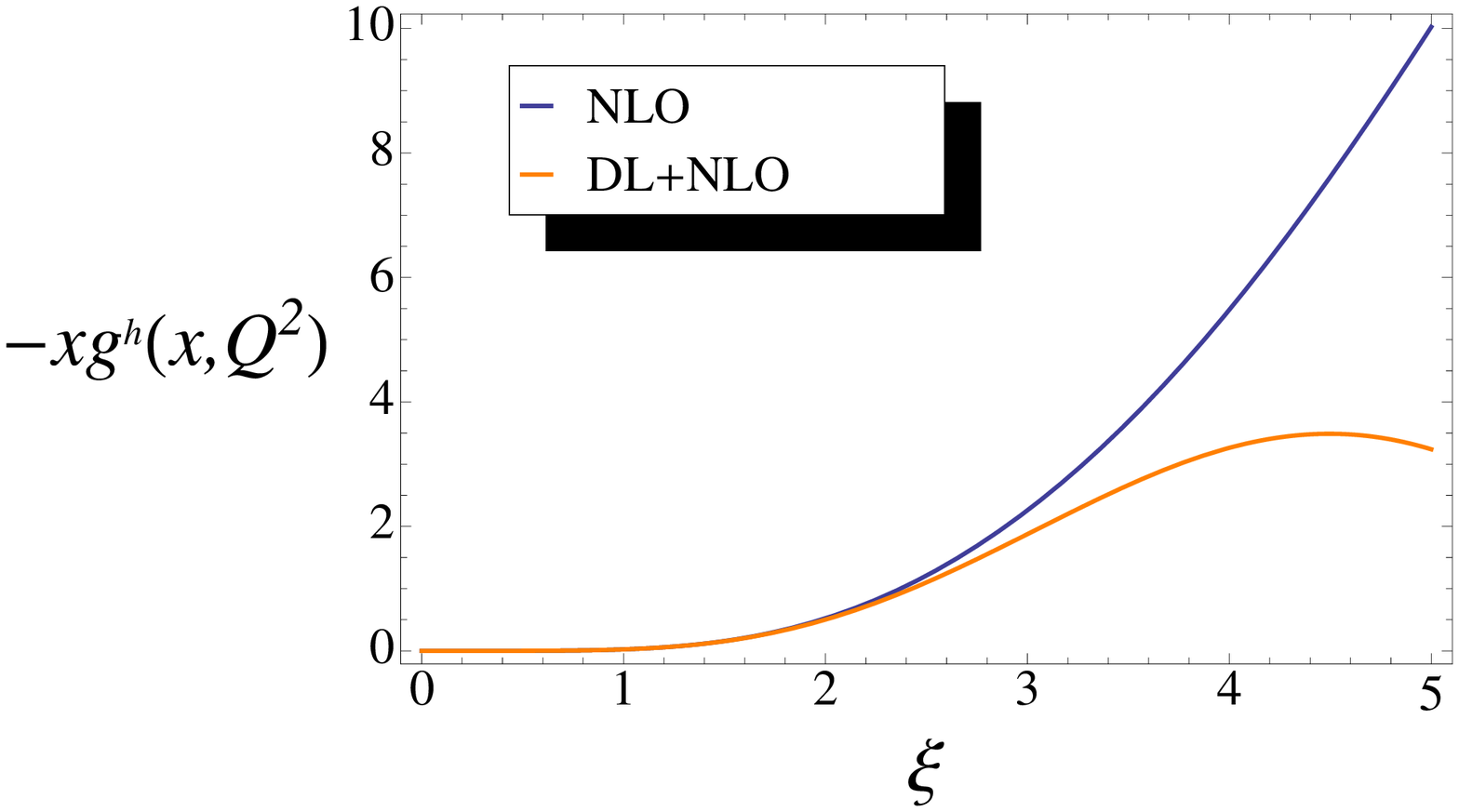}
\caption{\footnotesize{Contribution to the inclusive single hadron production cross section in $e^+ e^-$ annihilation due to the gluon
channel corrections in Eq.\ (\ref{gluonmultiplicity})} at NLO and with the inclusion of
the DLs at all orders in perturbation theory.
As in Fig.\ \ref{fig:coefficient}, the value $a_s(Q^2)=0.18$ was chosen in the calculation of
$C_g(z,a_s(Q^2))$, and the gluon FF was taken to be that obtained at $Q=14$ GeV in the LO global fit
of Ref.\ \cite{Albino:2005gd}.}
\label{fig:correction}
\end{minipage}
\end{center}
\end{figure}
In Fig.\ \ref{fig:coefficient}, we plot the NLO gluon coefficient function in Eq.\ (\ref{coeffNLO}),
together with the pure resummed result in Eq.\ (\ref{coeffDL}) and the combined DL+NLO 
result in Eq.\ (\ref{coeffDLpNLO}) as functions of $\xi=\ln(1/x)$. Note that the gluon coefficient function is negative.
We see that the DL+NLO result of Eq.\ (\ref{coeffDLpNLO})
interpolates well between its $O(a_s)$ FO result (Eq.\ (\ref{coeffNLO})) 
at large $x$ and the pure DL contribution
(Eq.\ (\ref{coeffDLpre})) at small $x$. 

We note that Eq.\ (\ref{coeffDLpNLO}) does not
show a DL divergent behaviour, or in fact any kind of divergent behaviour, for small $x$, in contrast to $C_g^{\rm{NLO}}$,
and this fact is illustrated in Fig.\ \ref{fig:coefficient} in the large $\xi$ region.
In fact, the quark coefficient function at NLO is also free of DLs and SLs (see e.g.\ Refs.\ \cite{Altarelli:1979kv,Baier:1979sp}).
In other words, there are no large logarithms at small $x$ at the NLO level after resummation of the DLs in the gluon coefficient function.
Thus it is possible that the LO analysis done in Refs.\ \cite{Albino:2005gd,Albino:2005gg} 
may now be extended to NLO, and for a similar range of data.
In Fig.\ (\ref{fig:correction}), we have estimated the contribution  
to the inclusive single hadron production cross section
coming from the gluon coefficient function in the $\overline{\rm MS}$
scheme, which is calculated according to the $x$-space convolution formula,
\be
g^h(x,Q^2)=\frac{1}{d\sigma_B}\frac{d\sigma^h}{dx}(x,Q^2)= \int_x^1 \frac{dx'}{x'}C_g(x',a_s(Q^2)) D_g^h\left(\frac{x}{x'},Q^2\right),
\label{gluonmultiplicity}
\ee
where all scales have been chosen equal and $D_g^h$ is the gluon FF.
The quantity $(1/d\sigma_B)(d\sigma^h/dx)$ has been evaluated with $C_g$ equal to
the NLO result in Eq.\ (\ref{coeffNLO}), and to the improved DL+NLO result in Eq.\ (\ref{coeffDLpNLO}), which
includes DLs at all orders.
The gluon FF has been chosen to be that obtained at $\sqrt{s}=14$ GeV 
in the global LO fit in Ref.\ \cite{Albino:2005gd}, in which all the DLs in the anomalous dimension
were resummed. As expected, this contribution is negative. Even if the correction seems
to be large, this does not imply a correction of the same amount in a NLO global fit,
because the gluon FF obtained in a LO analysis carries a large theoretical error.
Nevertheless, at the same time, this suggests that a sensible negative correction 
around and above the peak for the total hadron multiplicity is to be expected, 
thus necessitating a complete analysis using also a DL-resummed gluon coefficient function \cite{Albinofut}.

Finally, we note that, given that the SLs in the anomalous dimensions calculated in the MG scheme are already known \cite{Mueller:1983cq},
all that is needed to calculate the SLs in the anomalous dimensions calculated in the $\overline{\rm MS}$ scheme
is the relation between these two schemes at the DL level.
This relation is partially constrained by the fact that $C_g$ is now known in both schemes (Eqs.\ (\ref{gluoncoefff}) and (\ref{finalres})).
Thus, although a complete determination of the SLs in the anomalous dimensions calculated in the $\overline{\rm MS}$ scheme
is still not possible, some constraints may now be imposed on them which e.g.\ could be tested against FO results.
Further studies will be left to a future publication \cite{Albinofut}.

\section{Conclusions \label{conc}}

In this paper, we have computed and resummed all the DLs present in the gluon coefficient 
function. To resolve a mismatch between existing resummation formulae in the literature
\cite{Mueller:1982cq} and recent higher order corrections up to $O(\alpha_s^2)$
\cite{Rijken:1996vr,Rijken:1996npa,Rijken:1996ns,Mitov:2006ic,Mitov:2006wy,Blumlein:2006rr}, 
we have obtained the DL resummed result by applying dimensional regularization and then
subtracting the collinear singularities according to the $\overline{\rm MS}$
factorization scheme. The FO expansion of our new result agrees with the 
DLs of the full NNLO calculation.
We have seen that the resummation of the DLs in the anomalous dimensions and  
coefficient functions is sufficient to eliminate all large logarithms at NLO
that could destroy perturbation theory at very small $x$ values.
This could enable us to obtain observable quantities 
at NLO from data down to the same small values of $x$ as were previously treated
at LO in Refs.\ \cite{Albino:2005gd,Albino:2005gg}. We expect the impact of the DL-resummed gluon coefficient function
to be important, but not necessarily to modify drastically other results previously obtained, 
and we have pointed out the necessity of a complete DL+NLO analysis \cite{Albinofut}. 

\subsection*{Acknowledgments}

We thank J.\ Bl\"umlein, F.\ Caola and S.\ Forte for fruitful discussions.
This work was supported in part by the German Federal Ministry for Education
and Research BMBF through Grant No.\ 05~HT6GUA, by the German Research
Foundation DFG through the Collaborative Research Centre No.~676
{\it Particles, Strings and the Early Universe---The structure of Matter and
Space Time}, and by the Helmholtz Association HGF through the Helmholtz
Alliance Ha~101 {\it Physics at the Terascale}.
The work of A.V.K. was supported in part by DFG Grant No.\ INST 152/465--1,
Heisenberg-Landau Grant No.~5, and RFBR Grant No.\ 10--02--01259--a.

\begin{appendix}

\section{Factorization of multiple gluon emission}
\label{appendix}

Here we briefly sketch how the gluon emission probability factorizes in the 
kinematic region from which both soft and collinear DLs arise.
(``Factorize'' in this appendix refers to the vanishing of colour correlations.)
Here we recall only 
the features which are needed to understand the derivation of Eqs.\ (\ref{weight4xz}) 
and (\ref{expforprobdwD}) for the gluon emission
probability in $d=4$ and $d=4-2\epsilon$ dimensions, respectively, with $\epsilon <0$. 
For a more general and complete treatment about this factorization  
for infrared sensitive quantities in QCD jets, we refer the reader to 
Ref.\ \cite{Bassetto:1984ik}, of which the following is a summary.

We work in $d=4-2\epsilon$ dimensions.
We start by recalling that the QCD eikonal amplitude $\mathcal{E}_I(p,\epsilon(k))$ associated with a
soft gluon with momentum $k$ and polarization vector $\epsilon^{\mu}(k)$ that is emitted from a
parton $I$ ($I={\mathcal Q},$ if the parton is a quark, $\bar{\mathcal Q}$ if it is an antiquark, and $g$ if it is a gluon) of momentum $p$ is given 
in the case where $|\vec{k}| \ll |\vec{p}|$ by the result
\be
\mathcal{E}_I(p,k)= g\mu^\epsilon {\bf T}_I\, j_{\mu}(p,k)\epsilon^{*\mu}(k),
\label{eikampl}
\ee
where $j_{\mu}(p,k)=p_{\mu}/(p \cdot k)$.
${\bf T}_I$ are the colour charge
matrices which act on colour space only, and their matrix elements for the 
emission of a gluon of colour index $a$ are given by $({\bf T}_I)^{a}_{ij}\equiv t^{a}_{ij}$ 
(the colour matrix in the fundamental representation) if $I={\mathcal Q}$,
by $({\bf T}_I)^{a}_{ij}\equiv -t^{a}_{ij}$ if $I=\bar{\mathcal Q}$ and by 
$({\bf T}_I)^{a}_{bc}\equiv -if_{abc}$ (the colour matrix in the adjoint representation) 
if $I=g$. We remind that the colour-charge algebra
is given by:
\be
{\bf T}_I \cdot {\bf T}_J={\bf T}_J \cdot {\bf T}_I\quad \textrm{if $I\neq J$};\qquad 
{\bf T}_I^2=K_I\, {\bf{1}},
\label{colalg}
\ee
where $K_q=K_{\bar{q}}=C_F=(N_c^2-1)/2N_c$, $K_g=C_A=N_c$
and ${\bf{1}}$ is the identity colour charge operator in the fundamental 
and adjoint representations if $I={\mathcal Q},\bar{\mathcal Q}$ and $I=g$, respectively.
If we have many gluons in the final state, then radiation of gluons by other gluons can occur, and
the approximation in Eq.\ (\ref{eikampl}) can be applied iteratively when the momenta of these gluons are strongly ordered.

We now want to consider the more general physical case in which there are
$N$ soft gluons emitted with momenta
$k_\alpha$, where $\alpha=1,2,\dots,N$. We make the assumption that $|\vec{k_1}|\ll |\vec{k_2}| \ll \dots \ll |\vec{k_N}|\ll |\vec{p}|$.
Because we restrict our phase space to only one strongly ordered region, we treat each of the $N$ gluons as distinct when summing over the 
modulus squared amplitudes. 
Because the $|\vec{k_\alpha}|$ are negligible with respect to $|\vec{p}|$, we also have to add a recoiling
parton having momentum $\bar{p}$ with $\vec{\bar{p}}\approx -\vec{p}$ in order to respect momentum conservation.
Because the gluon with momentum $k_1$ is the softest gluon, 
the emission amplitude $\mathcal{E}_{\rm{tot}}(p,\bar{p},{\{}k_1,k_2,\dots,k_N{\}})$ for the whole process
can be calculated by calculating the emission amplitude $\mathcal{E}_{\rm{tot}}(p,\bar{p},{\{}k_2,k_3,\dots,k_N{\}})$
and then attaching the gluon with momentum $k_1$ and polarization 
$\epsilon(k_1)$ to each of the final state real partons.
The result is
\be
\mathcal{E}_{\rm{tot}}(p,\bar{p},{\{}k_1,k_2,\dots,k_N{\}})=\sum_{\alpha=2}^{N+2}\mathcal{E}_{I(\alpha)}(k_\alpha,k_1),
\label{colcoh}
\ee
where $I(\alpha)=g$ for $\alpha=1,2,\dots N$, $I(N+1)={\mathcal Q}$, $I(N+2)=\bar{\mathcal Q}$ and
where we define $k_{N+1}=p$ and $k_{N+2}=\bar{p}$. Now, performing the square modulus of this amplitude and summing
over the polarizations of the final state gluons, we determine the insertion operator for the soft
gluon to be
\be
{\bf I}(p,\bar{p},{\{}k_1,k_2,\dots,k_N{\}})=\sum_{\epsilon}\left|\mathcal{E}_{\rm{tot}}(p,\bar{p},{\{}k_\alpha{\}})\right|^2
=g^2\mu^{2\epsilon}\,\sum_{\alpha,\beta=2}^{N+2} {\bf T}_{I(\alpha)} \cdot {\bf T}_{I(\beta)}
\,j^{\mu}(k_\alpha,k_1)j^{\nu}(k_\beta,k_1) d_{\mu\nu}^{\,(n)}(k_1),
\label{sqtotamp}
\ee
where $d_{\mu\nu}^{\,(n)}$ is the on-shell polarization sum which, when parametrized by a gauge null 
vector $n^\mu$, reads
\be
d_{\mu\nu}^{\,(n)}(k)=-g_{\mu\nu}+\frac{k_\mu n_\nu+k_\nu n_\mu}{n \cdot k}.
\ee
Substituting this into Eq.\ (\ref{sqtotamp}), we find that the insertion operator can be written as
\be
{\bf I}(p,\bar{p},{\{}k_1,k_2,\dots,k_N{\}})=-g^2\mu^{2\epsilon}\, {\bf J}_\mu(p,\bar{p},{\{}k_1,k_2,\dots,k_N{\}}) {\bf J}^\mu(p,\bar{p},{\{}k_1,k_2,\dots,k_N{\}}),
\label{insop}
\ee
where
\be
{\bf J}^\mu(p,\bar{p},{\{}k_1,k_2,\dots,k_N{\}})=\sum_{\alpha=2}^{N+2}{\bf T}_{I(\alpha)} \left(\frac{k_\alpha^\mu}{k_\alpha \cdot k_1}
-\frac{n^\mu}{n \cdot k_1}\right).
\label{qcdcurrent}
\ee
We note that the quantity ${\bf J}^\mu(p,\bar{p},{\{}k_1,k_2,\dots,k_N{\}})$ is gauge 
invariant if we assume for the final state of the process without the gluon with momentum $k_1$
(described by $\mathcal{E}_{\rm{tot}}(p,\bar{p},{\{}k_2,k_3,\dots,k_N{\}})$)
the color singlet condition 
\be
\sum_{\alpha=2}^{N+2}{\bf T}_{I(\alpha)}=0,
\label{colcons}
\ee
which implies that the gauge vector $n^\mu$ can be chosen in a way that is convenient for calculations.
Due to colour correlations among the emitted gluons in Eq.\ (\ref{sqtotamp}),
in general we expect that the emission probability does not factorize.
Nevertheless, this does happen in the angular configuration from which the DL contribution arises, namely
the configuration in which $\vec{k_N}$ is approximately either parallel to $\vec{p}$ or anti-parallel to $\vec{p}$
(i.e.\ parallel to $\vec{\bar{p}}$),
and the gluons of momentum $k_\alpha$ for $\alpha=1,2,\dots,N-1$ have much smaller angles of emission
relative to the angle of emission of the gluon of momentum $k_N$,
and are approximately either all parallel or all anti-parallel to $\vec{k_N}$.
Because they then form a jet around either
$\vec{p}$ and $\vec{\bar{p}}$, i.e.\ either $k_\alpha^\mu \simeq \kappa_\alpha p^\mu$ or $k_\alpha^\mu \simeq \kappa_\alpha \bar{p}^\mu$ 
for $\alpha=1,2,\dots,N-1$,
where $\kappa_\alpha$ are some parameters that are independent of the spacetime index $\mu$.
For definiteness, we will assume that $\vec{k_N}$ is approximately parallel to $\vec{p}$ and then later multiply by a factor 2
to account for the case that $\vec{k_N}$ is approximately parallel to $\vec{\bar{p}}$. In this configuration, we find that the gluon 
with momentum $k_1$ factorizes and carries the total colour charge 
of the jet: We choose $n^\mu=\bar{p}$ in Eq.\ (\ref{qcdcurrent}) for convenience,
note that $k_\alpha^\mu/k_\alpha \cdot k_1=p^\mu/p \cdot k_1$ for $\alpha=2,3,\dots,N+1$,
and then use the gauge invariance condition in Eq.\ (\ref{colcons}) to write $\sum_{\alpha=2}^{N+1}{\bf T}_{I(\alpha)}=-{\bf T}_{I(N+2)}$.
If the initial state partons of momentum $p$ and $\bar{p}$ are gluons, then ${\bf T}_{I(N+2)}=T_g$.
If they are a quark and an antiquark, then, assuming that there are gluons already attached before the gluon of momentum $k_1$ is attached, 
these quarks must form a colour octet so that ${\bf T}_{I(N+2)}=T_g$ again.
Therefore, using the second result in Eq.\ (\ref{colalg}),
we find that the insertion operator in Eq.\ (\ref{insop}) now simply becomes
\be
{\bf I}(p,\bar{p},{\{}k_1,k_2,\dots,k_N{\}})=g^2\mu^{2\epsilon}\, K_g \, \frac{2 (p \cdot \bar{p})}{(p \cdot k_1) (\bar{p} \cdot k_1)}.
\label{eqforI}
\ee
Hence the single gluon emission probability for this angular configuration
factorizes according to
\be
d\sigma_N(p,\bar{p},k_1,k_2,\dots,k_N)=d\sigma_{N-1}(p,\bar{p},k_2,k_3,\dots,k_N)dw_g(k_1),
\label{singfacth}
\ee
where
\be
dw_I(k)=\,g^2\mu^{2\epsilon}\, K_I
\, \frac{2 (p \cdot \bar{p})}{(p \cdot k) (\bar{p} \cdot k)}\,
\frac{d^{d-1}k}{(2\pi)^{d-1}2k^0}
\label{eqfordwIinddim}
\ee
is the probability of emission of a single soft gluon of momentum 
$k$ from a parton $I$.
Here, $d\sigma_N$ is the $N$-gluon emission cross section and 
the last factor is the usual $d$-dimensional phase space measure.
In Eq.\ (\ref{singfacth}), all momenta $k_\alpha$ are fixed, where $\alpha=1,2,\dots,N$.
Note that the gluon with momentum $k_1$ is distinguished from the others only by the fact that it is the softest.
Thus, because the gluon with momentum $k_2$ is the softest one in the process described by $d\sigma_{N-1}(p,\bar{p},k_2,\dots,k_N)$,
by the same reasoning we have
\be
d\sigma_{N-1}(p,\bar{p},k_2,k_3,\dots,k_N)=d\sigma_{N-2}(p,\bar{p},k_3,k_4,\dots,k_N)dw_g(k_2).
\label{singfacth2}
\ee
Therefore, iterating this procedure, we find
\be
d\sigma_N(p,\bar{p},k_1,k_2,\ldots,k_N)=d\sigma_B(p,\bar{p})dw_g(k_1) dw_g(k_2) \ldots dw_g(k_{N-1}) dw_{I(N+2)}(k_N).
\label{dsigNp1fromdsigN2}
\ee
Note that, in the last step, we have used
$d\sigma_1(p,\bar{p},k_N)=d\sigma_B(p,\bar{p})dw_{I(N+2)}(k_N)$, where $dw_{I(N+2)} \neq dw_g$ necessarily because
$T_{I(N+2)}=T_q$ or $T_{\bar{q}}$ if the initial state partons are a quark and an antiquark:
they must be in a colour singlet state due to the fact that there are no gluons in the final state
before the gluon with momentum $k_N$ is attached.
Note that the same result in Eq.\ (\ref{dsigNp1fromdsigN2}) would have been obtained if we had used a different strong ordering for the
momenta $k_\alpha$ for $\alpha=1,2,\dots,N-1$.
This is a special case of the invariance under choice of strong ordering that is proved in Ref.\ \cite{Bassetto:1982ma}.

\end{appendix}

\bibliography{bibMSbarDLs}

\begin{thebibliography}{19}
\expandafter\ifx\csname natexlab\endcsname\relax\def\natexlab#1{#1}\fi
\expandafter\ifx\csname bibnamefont\endcsname\relax
  \def\bibnamefont#1{#1}\fi
\expandafter\ifx\csname bibfnamefont\endcsname\relax
  \def\bibfnamefont#1{#1}\fi
\expandafter\ifx\csname citenamefont\endcsname\relax
  \def\citenamefont#1{#1}\fi
\expandafter\ifx\csname url\endcsname\relax
  \def\url#1{\texttt{#1}}\fi
\expandafter\ifx\csname urlprefix\endcsname\relax\def\urlprefix{URL }\fi
\providecommand{\bibinfo}[2]{#2}
\providecommand{\eprint}[2][]{\url{#2}}

\bibitem[{\citenamefont{Mueller}(1981)}]{Mueller:1981ex}
\bibinfo{author}{\bibfnamefont{A.~H.} \bibnamefont{Mueller}},
  \bibinfo{journal}{Phys. Lett.} \textbf{\bibinfo{volume}{B104}},
  \bibinfo{pages}{161} (\bibinfo{year}{1981}).

\bibitem[{\citenamefont{Albino et~al.}(2006)\citenamefont{Albino, Kniehl,
  Kramer, and Ochs}}]{Albino:2005gd}
\bibinfo{author}{\bibfnamefont{S.}~\bibnamefont{Albino}},
  \bibinfo{author}{\bibfnamefont{B.~A.} \bibnamefont{Kniehl}},
  \bibinfo{author}{\bibfnamefont{G.}~\bibnamefont{Kramer}}, \bibnamefont{and}
  \bibinfo{author}{\bibfnamefont{W.}~\bibnamefont{Ochs}},
  \bibinfo{journal}{Phys. Rev.} \textbf{\bibinfo{volume}{D73}},
  \bibinfo{pages}{054020} (\bibinfo{year}{2006}), \eprint{hep-ph/0510319}.

\bibitem[{\citenamefont{Albino et~al.}(2005)\citenamefont{Albino, Kniehl,
  Kramer, and Ochs}}]{Albino:2005gg}
\bibinfo{author}{\bibfnamefont{S.}~\bibnamefont{Albino}},
  \bibinfo{author}{\bibfnamefont{B.~A.} \bibnamefont{Kniehl}},
  \bibinfo{author}{\bibfnamefont{G.}~\bibnamefont{Kramer}}, \bibnamefont{and}
  \bibinfo{author}{\bibfnamefont{W.}~\bibnamefont{Ochs}},
  \bibinfo{journal}{Phys. Rev. Lett.} \textbf{\bibinfo{volume}{95}},
  \bibinfo{pages}{232002} (\bibinfo{year}{2005}), \eprint{hep-ph/0503170}.

\bibitem[{\citenamefont{Mueller}(1983{\natexlab{a}})}]{Mueller:1982cq}
\bibinfo{author}{\bibfnamefont{A.~H.} \bibnamefont{Mueller}},
  \bibinfo{journal}{Nucl. Phys.} \textbf{\bibinfo{volume}{B213}},
  \bibinfo{pages}{85} (\bibinfo{year}{1983}{\natexlab{a}}).

\bibitem[{\citenamefont{Rijken and van Neerven}(1996)}]{Rijken:1996vr}
\bibinfo{author}{\bibfnamefont{P.~J.} \bibnamefont{Rijken}} \bibnamefont{and}
  \bibinfo{author}{\bibfnamefont{W.~L.} \bibnamefont{van Neerven}},
  \bibinfo{journal}{Phys. Lett.} \textbf{\bibinfo{volume}{B386}},
  \bibinfo{pages}{422} (\bibinfo{year}{1996}), \eprint{hep-ph/9604436}.

\bibitem[{\citenamefont{Rijken and van
  Neerven}(1997{\natexlab{a}})}]{Rijken:1996npa}
\bibinfo{author}{\bibfnamefont{P.~J.} \bibnamefont{Rijken}} \bibnamefont{and}
  \bibinfo{author}{\bibfnamefont{W.~L.} \bibnamefont{van Neerven}},
  \bibinfo{journal}{Phys. Lett.} \textbf{\bibinfo{volume}{B392}},
  \bibinfo{pages}{207} (\bibinfo{year}{1997}{\natexlab{a}}),
  \eprint{hep-ph/9609379}.

\bibitem[{\citenamefont{Rijken and van
  Neerven}(1997{\natexlab{b}})}]{Rijken:1996ns}
\bibinfo{author}{\bibfnamefont{P.~J.} \bibnamefont{Rijken}} \bibnamefont{and}
  \bibinfo{author}{\bibfnamefont{W.~L.} \bibnamefont{van Neerven}},
  \bibinfo{journal}{Nucl. Phys.} \textbf{\bibinfo{volume}{B487}},
  \bibinfo{pages}{233} (\bibinfo{year}{1997}{\natexlab{b}}),
  \eprint{hep-ph/9609377}.

\bibitem[{\citenamefont{Mitov and Moch}(2006)}]{Mitov:2006wy}
\bibinfo{author}{\bibfnamefont{A.}~\bibnamefont{Mitov}} \bibnamefont{and}
  \bibinfo{author}{\bibfnamefont{S.}~\bibnamefont{Moch}},
  \bibinfo{journal}{Nucl. Phys.} \textbf{\bibinfo{volume}{B751}},
  \bibinfo{pages}{18} (\bibinfo{year}{2006}), \eprint{hep-ph/0604160}.

\bibitem[{\citenamefont{Blumlein and Ravindran}(2006)}]{Blumlein:2006rr}
\bibinfo{author}{\bibfnamefont{J.}~\bibnamefont{Blumlein}} \bibnamefont{and}
  \bibinfo{author}{\bibfnamefont{V.}~\bibnamefont{Ravindran}},
  \bibinfo{journal}{Nucl. Phys.} \textbf{\bibinfo{volume}{B749}},
  \bibinfo{pages}{1} (\bibinfo{year}{2006}), \eprint{hep-ph/0604019}.

\bibitem[{\citenamefont{Mitov et~al.}(2006)\citenamefont{Mitov, Moch, and
  Vogt}}]{Mitov:2006ic}
\bibinfo{author}{\bibfnamefont{A.}~\bibnamefont{Mitov}},
  \bibinfo{author}{\bibfnamefont{S.}~\bibnamefont{Moch}}, \bibnamefont{and}
  \bibinfo{author}{\bibfnamefont{A.}~\bibnamefont{Vogt}},
  \bibinfo{journal}{Phys. Lett.} \textbf{\bibinfo{volume}{B638}},
  \bibinfo{pages}{61} (\bibinfo{year}{2006}), \eprint{hep-ph/0604053}.

\bibitem[{\citenamefont{Bassetto et~al.}(1982)\citenamefont{Bassetto,
  Ciafaloni, Marchesini, and Mueller}}]{Bassetto:1982ma}
\bibinfo{author}{\bibfnamefont{A.}~\bibnamefont{Bassetto}},
  \bibinfo{author}{\bibfnamefont{M.}~\bibnamefont{Ciafaloni}},
  \bibinfo{author}{\bibfnamefont{G.}~\bibnamefont{Marchesini}},
  \bibnamefont{and} \bibinfo{author}{\bibfnamefont{A.~H.}
  \bibnamefont{Mueller}}, \bibinfo{journal}{Nucl. Phys.}
  \textbf{\bibinfo{volume}{B207}}, \bibinfo{pages}{189} (\bibinfo{year}{1982}).

\bibitem[{\citenamefont{Ellis et~al.}(1979)\citenamefont{Ellis, Georgi,
  Machacek, Politzer, and Ross}}]{Ellis:1978ty}
\bibinfo{author}{\bibfnamefont{R.~K.} \bibnamefont{Ellis}},
  \bibinfo{author}{\bibfnamefont{H.}~\bibnamefont{Georgi}},
  \bibinfo{author}{\bibfnamefont{M.}~\bibnamefont{Machacek}},
  \bibinfo{author}{\bibfnamefont{H.~D.} \bibnamefont{Politzer}},
  \bibnamefont{and} \bibinfo{author}{\bibfnamefont{G.~G.} \bibnamefont{Ross}},
  \bibinfo{journal}{Nucl. Phys.} \textbf{\bibinfo{volume}{B152}},
  \bibinfo{pages}{285} (\bibinfo{year}{1979}).

\bibitem[{\citenamefont{Curci et~al.}(1980)\citenamefont{Curci, Furmanski, and
  Petronzio}}]{Curci:1980uw}
\bibinfo{author}{\bibfnamefont{G.}~\bibnamefont{Curci}},
  \bibinfo{author}{\bibfnamefont{W.}~\bibnamefont{Furmanski}},
  \bibnamefont{and}
  \bibinfo{author}{\bibfnamefont{R.}~\bibnamefont{Petronzio}},
  \bibinfo{journal}{Nucl. Phys.} \textbf{\bibinfo{volume}{B175}},
  \bibinfo{pages}{27} (\bibinfo{year}{1980}).

\bibitem[{\citenamefont{Mueller}(1983{\natexlab{b}})}]{Mueller:1983js}
\bibinfo{author}{\bibfnamefont{A.~H.} \bibnamefont{Mueller}},
  \bibinfo{journal}{Nucl. Phys.} \textbf{\bibinfo{volume}{B228}},
  \bibinfo{pages}{351} (\bibinfo{year}{1983}{\natexlab{b}}).

\bibitem[{\citenamefont{Altarelli et~al.}(1979)\citenamefont{Altarelli, Ellis,
  Martinelli, and Pi}}]{Altarelli:1979kv}
\bibinfo{author}{\bibfnamefont{G.}~\bibnamefont{Altarelli}},
  \bibinfo{author}{\bibfnamefont{R.~K.} \bibnamefont{Ellis}},
  \bibinfo{author}{\bibfnamefont{G.}~\bibnamefont{Martinelli}},
  \bibnamefont{and} \bibinfo{author}{\bibfnamefont{S.-Y.} \bibnamefont{Pi}},
  \bibinfo{journal}{Nucl. Phys.} \textbf{\bibinfo{volume}{B160}},
  \bibinfo{pages}{301} (\bibinfo{year}{1979}).

\bibitem[{\citenamefont{Baier and Fey}(1979)}]{Baier:1979sp}
\bibinfo{author}{\bibfnamefont{R.}~\bibnamefont{Baier}} \bibnamefont{and}
  \bibinfo{author}{\bibfnamefont{K.}~\bibnamefont{Fey}}, \bibinfo{journal}{Z.
  Phys.} \textbf{\bibinfo{volume}{C2}}, \bibinfo{pages}{339}
  (\bibinfo{year}{1979}).

\bibitem[{\citenamefont{Albino et~al.}(in preparation)\citenamefont{Albino,
  Bolzoni, Kniehl, and Kotikov}}]{Albinofut}
\bibinfo{author}{\bibfnamefont{S.}~\bibnamefont{Albino}},
  \bibinfo{author}{\bibfnamefont{P.}~\bibnamefont{Bolzoni}},
  \bibinfo{author}{\bibfnamefont{B.}~\bibnamefont{Kniehl}}, \bibnamefont{and}
  \bibinfo{author}{\bibfnamefont{A.}~\bibnamefont{Kotikov}} (\bibinfo{year}{in
  preparation}).

\bibitem[{\citenamefont{Mueller}(1984)}]{Mueller:1983cq}
\bibinfo{author}{\bibfnamefont{A.~H.} \bibnamefont{Mueller}},
  \bibinfo{journal}{Nucl. Phys.} \textbf{\bibinfo{volume}{B241}},
  \bibinfo{pages}{141} (\bibinfo{year}{1984}).

\bibitem[{\citenamefont{Bassetto et~al.}(1983)\citenamefont{Bassetto,
  Ciafaloni, and Marchesini}}]{Bassetto:1984ik}
\bibinfo{author}{\bibfnamefont{A.}~\bibnamefont{Bassetto}},
  \bibinfo{author}{\bibfnamefont{M.}~\bibnamefont{Ciafaloni}},
  \bibnamefont{and}
  \bibinfo{author}{\bibfnamefont{G.}~\bibnamefont{Marchesini}},
  \bibinfo{journal}{Phys. Rept.} \textbf{\bibinfo{volume}{100}},
  \bibinfo{pages}{201} (\bibinfo{year}{1983}).

\end{thebibliography}

\end{document}